\documentstyle[twocolumn,aps,prb,epsf]{revtex}
\begin{document}
\preprint{ICAMS 18}
\draft
\title{Anderson transition and thermal effects on electron states in amorphous silicon} 
\author{D. A. Drabold}\protect
\address{Department of Physics and Astronomy, Condensed
Matter\\
and Surface Science Program, Ohio University, Athens, Ohio 45701-2979, USA\\
http://www.phy.ohiou.edu/$^\sim$drabold/drabold.html}
\maketitle
\vskip .7cm
\noindent{\bf Abstract}
I discuss the properties of electron states in amorphous Si based on large scale calculations
with realistic several thousand atom models. A relatively simple model for the localized
to extended (Anderson) transition is reviewed. Then, the effect of thermal disorder
on localized electron states is
considered. It is found that under readily accessible
conditions, localized (midgap or band tail) states
and their conjugate energies may
fluctuate dramatically. The possible importance of non-adiabatic 
atomic dynamics to doped or photo-excited 
systems is briefly discussed. 

\vskip .5truecm
\noindent{\bf 1. Introduction}

There have been a very large number of studies on electron states in amorphous
materials, and especially amorphous silicon~[1]. These computations have ranged from
simplified tight-binding approaches to first principles methods using 
density functional theory in the
local spin density approximation (LSDA)~[2]. 
A difficulty with many calculations of electron states in amorphous Si is the usual
challenge of obtaining a realistic starting model. Amorphous Si is well known to be
almost entirely four-coordinated with fewer than one atom in 10$^3$ being non-ideally
coordinated. However, few if any models fabricated entirely with 
molecular dynamics (MD) (including first
principles MD) have a defect concentration at all consistent with experiment. In
particular, models obtained by a quench from the melt are particularly unrealistic,
often with 10\% or more coordination defects in a cell of order 100 atoms. Such a model
does not posess a true optical gap, but rather just a deep minimum in the density of
states near the Fermi level. This result is inconsistent with electron
spin resonance (ESR) measurements and
most other estimates of defect concentration
in the material. Such models are probably best interpreted as poor metals rather
than semiconductors, since there is so much banding between impurities that the defect
states are fully extended through the supercell. The origin of this trouble is easy
to ascertain: the liquid (a predominantly sixfold {\it metal})
is topologically quite distinct from amorphous (or cystalline) Si, and any rapid
quench is certain to freeze in too many remnants of the liquid (such as overcoordination defects,
which are more common than dangling bonds in most published MD simulations~[2]). 
Because of the qualitative difference in liquid and amorphous
topology, it is perhaps
not surprising that a-Si cannot be formed experimentally from a quench from the 
liquid.
We have shown elsewhere~[2] that a proper description of defect 
wave functions (with localization
consistent with estimates from ESR) requires both a realistic model (with a very small
defect concentration less that 1 \% ) and LSDA, to estimate the spin localization
of an isolated defect. Fortunately realistic models (meaning in uniform agreement with
structural, vibrational and electronic/optical experiments) do exist, but are made
using Wooten-Weaire-Winer~[3] methods. An excellent set of these models 
(with 216, 512 and 4096 atoms)
was
produced by Djordjevic, Thorpe and Wooten~[4]; we use these in the studies presented here.

It is
now established that the midgap defect state is due to dangling bonds. 
The floating bond conjecture (that five-fold
coordinated atoms play an important role in a-Si) has not been
disproven, though any such states are shifted well toward the
conduction band edge~[2] and are not nearly
so localized as the sp$^3$ dangling bonds. If such floating bond defects exist, they might form resonances 
with the conduction tail states, which could have interesting implications for 
transport in n-type materials.

Less progress has been made on band tail states, because a good description
of these requires models with large numbers (several hundred to thousands) of atoms. To obtain
a qualitative understanding of the Anderson transition in a three-dimensional network
we used a 4096 atom supercell model~[4] 
(a cube of amorphous Si about 4 nm on a side). The band tail states have
structure on a length
scale several times the nearest neighbor distance (depending on the energy), and an 
enormous model (by the standards of electronic structure calculations) is essential
to disentangle the ``simple" features of the band tail tates in amorphous Si. In section
3 we describe the properties of these states.

Finally we discuss the electronic effects 
due to thermal disorder superposd on a topologically
disordered network~[5]. We observe from a straightforward thermal simulation
of an amorphous Si network that well-localized states can have their energies modulated
by {\it tenths} of an eV (interestingly, an order of magnitude larger than the thermal energy 
itself). Also, the
states themselves can fluctuate spatially as we describe in section 4 below. This
feature is of importance to transport, optical and doping studies.

\vskip 0.5truecm
\noindent{\bf 2. Methodology}

For the calculations of electronic tail states in the 4096 atom model, we
used a tight-binding Hamiltonian and a sophisticated Lanzcos scheme to compute
about 500 states near the optical gap~[6] (the dimension of the Hamiltonian matrix
is 16384). A technical point of some importance is that a ``simple" Lanczos scheme 
designed for computing
states near a selected energy, such as the folded spectrum method (computing
states from a shifted and squared  {\bf H} matrix) is not adequate
to compute more than about 20 states here because the spectrum becomes
exceedingly dense for energies out of the middle of the gap. What
is effective is a ``shift and invert" strategy, which spreads the spectrum out
effectively, enabling the reliable calculation of many more states. Thus,
one computes the matrix ${\bf{G}}(E)~ = ~(E{\bf 1} - {\bf H})^{-1}$; and performs an elementary
Lanczos calculation on the matrix ${\bf{G}}$ (to obtain spectral information
near E; ${\bf H}$ is the Hamiltonian matrix, ${\bf 1}$ is
the unit matrix). It is
amusing that this method was apparently pioneered by engineers~[7]
instead of physicists, since the object
${\bf{G}}$ is so familiar to workers in condensed matter 
physics (it is the electronic Greens
matrix!). It is not {\it a priori} obvious that the expense of inverting 
$(E {\bf 1} - {\bf H}$) is worthwhile, but because ${\bf H}$ is sparse this is not prohibitively
expensive, and the transformed spectrum is well conditioned for Lanczos (the
eigenvalues of ${\bf{G}}(E)$ are well separated near $E$ even if the spectrum of
{\bf H} is dense near $E$).

For dynamical simulation, we use the methods of Sankey and coworkers~[8] ``Fireball96",
which is discussed in the context of amorphous Si elsewhere~[5].

\vskip 0.5truecm
\noindent{\bf 3. Model of the Anderson transition in amorphous Si}

I have discussed the details of the computation of gap and band tail states
in the 4096 atom model of amorphous Si elsewhere~[6]; here I dwell upon the
``physics" of the calculation. In a nutshell, we found that (1) states near
the middle of the gap were exponentially localized (as one would expect); (2)
for energies shifted from midgap toward the valence tail, there was a tendency
to find system eigenstates with rather similar charge densities, and consisting
approximately of two separated ``blobs" of charge; (3) deeper into the valence tail we
observed that the states still were discernibly ``built" from clumps or clusters
of charge, with a clear separation between these clumps; (4) deep into the valence
tail we saw states that were well extended, essentially uniformly, through the cell.
See Figs. 1 and 2 in Ref.~[6]; (5) {\it The states are very nonlocal} away from midgap,
in contradiction with the defect pool model; it is clear that this must happen since the
states {\it in the bands} are extended; this computation shows that states are really
well localized only within of order $~\sim 0.6-0.7 eV$ from midgap.

The preceding discussion suggests that an alternate basis or representation may
be worthwhile to describe the electron states of a-Si, especially near the
gap. As usual in quantum mechanics, one is at liberty to select any complete
(or at least approximately complete) set to represent the quantum mechanical
states. The choice is normally made on grounds of efficiency: basis functions
are selected which capture much of the character of the eigenfunctions, so that
only a small number of basis functions is required to faithfully represent the
eigenfunctions. The classic example is the method of orthogonalized plane 
waves in band theory~[9]. Likewise here, since the eigenvectors in the gap and near
the tails are clearly composed of ``lumps" or clusters of charge, a basis consisting of
such cluster states is appealing.

I take the view that the cluster states are localized eigenfunctions which arise from
idealized isolated defects. That is, one can imagine all possible distortions of
a continuous random network. Select one such distortion (say 
a strained bond angle
removed from the distribution of bond angles in the system). Suppose further that the 
strain is large enough to make the state bound (eg. localized) 
with exponential tail. Then
the state has both a well defined energy and well defined extent in real space. Such a 
situation certainly exists in a-Si provided that the defect is really isolated (in space) from
resonant defects. From this point of view, the highly localized states near the middle of 
the gap arising from single defects in large cells are examples of 
especially spatially compact cluster states. As one
considers energies closer to the valence and conduction bands (in the band tail regions), it
becomes difficult to compute the cluster states exactly, since the density of states increases
(which implies that the number of cluster states is also greater) so that the probablity 
is large
that
the tails of resonant defects overlap. If there is such an overlap between resonant defects
(clusters), simple
quantum mechanics guarantees that 
eigenstates will consist of mixtures of cluster states with formation
of ``bonding and antibonding" 
levels. 
Another way of saying it is that we are in the curious position of: ``Given the
eigenvectors, what is the most localized basis with smallest number of elements which can be superposed
to reproduce the eigenstates?" Such cluster states have much in common with our recent work
on generalized Wannier functions in amorphous materials~[10].

Our qualitative picture of the local to extended transition in a-Si, 
based on these
calculations is the following:
As severe distortions are rare,
clusters stemming from such distortions
are probably isolated from each other and if isolated, are 
localized 
energy eigenstates. For less severe distortion, the
probability of occurrence increases, and the size of associated clusters
is also larger. Then the chance of finding another cluster of similar
energy in the neighborhood 
increases. As the distortion becomes
less severe, eigenstates will
consist of mixtures of two, three, or more clusters. At
some point, clusters can always find ``overlapping energy partners"
and they
mix together to enable electronic connectivity (extended eigenfunctions). This state of affairs
can be identified with the ``mobility edge".

With the preceding introduction, it is natural to write down a simple 
``theory" utilizing these
ideas. A reasonable form for this model is:
\begin{eqnarray}
{\hat{H}} = \sum_\alpha E_\alpha |\alpha \rangle \langle\alpha| +
\sum_{\alpha \beta (\alpha \neq \beta)}|\alpha 
\rangle\xi_{\alpha\beta} \langle\beta|
\end{eqnarray}
where, basis state $|\alpha \rangle$ is a localized cluster
({\it which may involve many atoms}).
In this representation, the
basis functions are
localized energy eigenstates with
eigenvalue $E_\alpha$ of the Hamiltonian in the absence of other
defects with which mixing occurs. 
$E_\alpha$ is determined by the distortion.
In real amorphous solids,
the cluster states may significantly overlap. $\xi_{\alpha\beta}$
represents the coupling between cluster $| \alpha \rangle$ and
$| \beta \rangle$. In the spirit of H\"uckel theory,
we can take:
$\xi_{\alpha \beta} \sim (E_\alpha+E_\beta) S_{\alpha \beta}/2$ for
$S_{\alpha \beta} =  \langle\alpha|\beta \rangle$. Then, in first order perturbation 
theory, the formation of eigenstates of ${\bf H}$ from these clusters
becomes obvious; the first order correction to the zeroth-order
(cluster) state $|\alpha \rangle$ is $\sum_{\beta \neq \alpha} 
\Gamma_{\alpha \beta} |\beta \rangle$, where
$ \Gamma_{\alpha \beta}=(E_\alpha 
+E_\beta)S_{\beta \alpha}/2(E_\alpha-E_\beta)$. The strong mixing for
small energy denominators ($E_\alpha \approx E_\beta$), and the role of the
overlap is indicated. For reasons that are apparent, we have called this 
the ``resonant cluster
proliferation" model of the Anderson transition~[6].

Since the cluster basis states $| \alpha \rangle$ are not precisely known, no calculations
have yet been attempted in this representation. It is however easy to imagine a calculation
in which model cluster states are introduced with characteristic decay lengths; and the role
of the amorphous net is largely encoded into $\xi_{\alpha \beta}$. Even in the absence of
new explicit calculations, I assume that the ``resonant cluster" model is useful to
understanding the Anderson transition, and the role of thermal disorder in a topologically
disordered network, as I describe next.

\vskip .5cm
\noindent{\bf 4. Thermal modulation of electron states and energies}

There is abundant experimental evidence that lattice vibrations play an important role in the
dynamics of electrons in amorphous materials. Among other examples, Cohen and 
coworkers~[11] observed a pronounced temperature
dependence of the Urbach tails in a-Si:H (the conduction tails showing a very strong linear variation
in decay parameter with temperature). Of course electrical conductivity is well known to be
very temperature dependent, and usually has multiple distinct regimes according
to different conduction
mechanisms~[12]. From this point of view it is 
unsurprising that the electron energies and states
can be very time and temperature dependent. This point has been 
independently recognized by Arkhipov and Adriaennsens~[13]
in their studies of carrier transport.

We study thermal modulation of the eigenvalues and eigenvectors by 
tracking their (adiabatic) time development over the course of a 
few picosecond ($10^{-12}$s) simulation. As illustrated in Fig. 1 in
Ref.~[5], there is strong time dependence
of the LDA eigenvalues in
the vicinity of the optical gap. The Fermi level is near the middle of the
gap and several states near the Fermi level are appropriately described
as band tail states. These are much like the states which would be responsible
for conduction in doped a-Si:H. 
As in earlier work~[14] there is a roughly linear relation between root mean square
(rms) temporal fluctuation and temperature.
As expected, the higher
temperature simulation leads to larger excursions in the positions of the
energy eigenvalues. Note for 300K that the 
Lowest Unoccupied Molecular Orbital (LUMO) fluctuates in time by about
$\sim$0.3eV, {\it much} larger than thermal energies ($\sim$10 meV). 
States deeper into either the valence or conduction bands 
show progressively less thermal modulation because they are less localized 
(we
have noted~[15] a very strong correlation between the
rms fluctuation in the energy eigenvalues due to thermal disorder and the
inverse participation ratio, a simple measure of localization in the
T=0 model). The localization ``amplifies" the electron-phonon coupling. 
Also, the conduction states fluctuate more
than the valence states (suggesting that the conduction tails are
more sensitive to thermal disorder than the valence tails which originate
primarily from structural disorder), in pleasing agreement
with total yield photoemission experiments~[11] and earlier
theory work~[14]. 

Elsewhere ~[5], we have published color figures depicting the fluctuations in
electron states near the gap for the system and dynamics described above. There
are very substantial changes in the LUMO state in particular;
There is a clear tendency for the LUMO state to alternately
``accumulate" on a strained part of the network, sometimes becoming
localized, but also occasionally developing 
a substantially more extended
``stringlike" character. 
These are not the only two recognizable structures, but
recur most frequently. 
The time between ``characters" is not predictable,
though it is of order tens to hundreds of fs. We have posted an animation of
this state on the world wide web~[16].

We have argued in the preceding section that structural disorder in a-Si
gives rise to localized states with energies in the band tails. These
system eigenstates can involve 
{\it many} atoms and can have a Byzantine~[17] structure.
The ``simple physics" of this study is that the
strong (compared to midband electrons) electron-phonon coupling for localized band tail 
states {\it is sufficient}
to cause strongly time/temperature dependent quantum mechanical mixing of cluster states when
the thermal disorder is ``just right" to make their energies nearly 
degenerate provided that they have some overlap in real space.
Strong mixing of course implies less localization and thus better
prospects, at least while the more extended state survives, for conductivity
and optical transitions.
This work shows that transport and optical calculations
based only on T=0 results can be quite misleading.
Mott~[12] and others have made fundamental 
contributions to the theory of transport in a noncrystalline medium; for
example,variable range hopping.
In the kind of simulation we present here,
I can estimate the conductivity, including its temperature and frequency
dependence {\it directly} from the electronic states through an appropriate
thermal average of the Kubo formula~[18].
It is also a complement to the phenomenological
models of transport~[13,19]. In the latter work, transport is
modeled as a hopping between localized tail states. My work can be
viewed as an explanation of the precise nature of the states among
which electrons are hopping (the very complicated states of
Ref. 6). 
The waiting time between hops must be related to
the time between eigenvalue "close encounters" near the Fermi level.
It also points at an atomistic level to the dynamics of bandtail defects
and their kinetics.

The consequences of this work can be stated
another way. If $|i \rangle$ ($|f \rangle$) are initial (final) 
electronic states
with energy $E_i$ ($E_f$), then 
for an electronic transition in a-Si, a Fermi golden rule argument leads
quicky to the conclusion~[12] that the transition rate is
proportional 
to $|\langle i | \hat{T} | f \rangle|^2 \delta(E_f-E_i-\hbar \omega)$,
where $\hat{T}$ is a perturbation inducing the 
transition (to first approximation
a momentum operator) and $\omega$ is the frequency of an external probe. 
Both the energies in the $\delta$ function and the transition matrix elements
are affected by the instantaneous details of the structural disorder, and
as such transition probabilities
are also strongly dependent on
the time and temperature. The consequences of this to transport
are under investigation; the discussion here is based upon first-order
time dependent perturbation theory, which for the very strong electron-phonon
coupling we discuss, could be inadequate.

\vskip .5cm
\noindent{\bf 5.  Heuristics}

Earlier work has shown that it is useul to link the thermal fluctuation of
the LDA energy eigenvalues near the band tails to the extent (localization) of the band tails
in amorphous Si (as separately measured in total yield photoemission experiments).
Here I give a simple model which makes this connection between fluctuating
electron energies and lattice vibrations a bit more explicit.

Consider a particular electronic eigenvalue, $\lambda_n$, say in one of the
band tails in a-Si. To estimate the sensitivity of $\lambda_n$ to a coordinate
distortion (supposedly thermally induced), we can use the 
Hellmann-Feynman theorem~[20],
which gives $\partial \lambda_n/\partial {\bf R_\alpha}~=~\langle \psi_n |
\partial {\bf H}/\partial {\bf R_\alpha} | \psi_n \rangle$ (for this to be valid,
I must assume that the basis is fixed (not
moving with the atoms) and that the $|\psi_n \rangle$ are
exact eigenvectors of ${\bf{H}}$;
see Ref.~[8] for a more general case). Then clearly for small distortions ${\bf R_\alpha}$, we
have $\delta \lambda_n ~\sim~ \sum_\alpha \langle \psi_n |
\partial {\bf{H}}/\partial {\bf R_\alpha} | \psi_n \rangle \delta {\bf R_\alpha}$. Here, {\bf R}
is the $3N$ vector of displacements for all of the atomic coordinates from equilibrium.
If the displacements $\delta {\bf R_\alpha} (t)$ arise from classical vibrations, then
one can also write $\delta {\bf R_\alpha}(t)~=~\sum_{\omega}~A(T,\omega)~
\cos[ \omega t + \phi_\omega] \chi_\alpha (\omega)$, where $\omega$ indexes the normal mode
frequencies, $A(T,\omega)$ is the temperature dependent amplitude of the mode with 
frequency $\omega$, $\phi_\omega$ is an arbitrary phase, and $\chi_\alpha(\omega)$ is a 
normal mode with frequency $\omega$ and vibrational displacment
index $\alpha$. Using a temperature dependent amplitude
$A^2(T,\omega)~=~ k_B T/2M\omega^2$, it is
easy to see that the trajectory (long time) average of the expression for $\delta 
\lambda_n^2 $ is: 
$\langle \delta \lambda_n^2 \rangle ~\sim~ (k_B T/4M) 
\sum_\omega (\Xi(\omega)/
\omega)^2$, where $\Xi(\omega)~=~[\sum_\alpha \langle \psi_n |
\partial {\bf H}/\partial 
{\bf R_\alpha} | \psi_n \rangle \chi_\alpha(\omega)]$. It would 
be straightforward for a particular collection of 
vibrational states to include
the correct Bose terms to obtain a result valid at low temperatures $T < \Theta_D$
(for $\Theta_D$ a salient Debye temperature). 
These expressions give a transparent
expression for the thermally induced
electron modulation as driven by the
lattice-electron coupling $\Xi(\omega)$. If we follow Ref.~[14] and
make the coarse approximation of equating $\langle \delta 
\lambda_n^2\rangle^{1/2}$ to the Urbach decay parameter~[11], then
this model would predict a square root dependence of the decay parameter
with temperature, perhaps not easily distinguished from the linear
dependence reported in Ref.~[11].

It is easy to see why there is a correlation between an
eigenstate's localization
(as measured for example by inverse participation ratio)
and its thermal rms fluctuation~[15]. Note that $\Xi(\omega)$ will in general consist of a sum
of many terms (for different $\alpha$); the individual terms have no preferred sign,
so that adding a large number of terms of comparable magnitude will lead to cancellation
and a small sum. On the other hand, if only a very small number of terms are nonzero
(as for the case when the electron state $|\psi_n \rangle$ is well localized), then there
will be less cancellation and a larger contribution to the sum (and therefore to the
fluctuation of the eigenvalue).
This model is limited in many
ways (it is classical, as all the modes are equally
populated from classical equipartition),
it is obviously strictly harmonic and
we are assuming no electronic level crossings or other departures
from adiabatic (Born-Oppenheimer) dynamics~[5]), but captures some of the right
temperature dependent electronic effects observed.

The thermal effects on the electron {\it states} are more difficult to calculate 
with this approach, mostly because of problems with degeneracy
if one tries to use perturbation theory. Still, the underlying ``physics" is 
quite simple, just the effects of resonant mixing for close approach (in energy) of
eigenvalues originating in cluster states with some overlap in space. In the language
of the model Hamiltonian above, this means that there can be large mixing
of cluster basis states when the energy denominator becomes small; here the
thermal fluctations can induce small energy denominators (and therefore mixing).
Of course a mixed state involves more cluster states than a pure cluster
state, so that it is more spatially extended. 

\vskip .5cm
\noindent{\bf 5. Post adiabatic atomic dynamics}

The calculations and discussion of the preceding sections has been based on the
adiabatic approximation (that the electrons are always in their instantaneous ground
state determined by the external potential (eg the atomic nuclei), and that the
forces ``felt" by the nuclei arose from the instantaneous electronic structure of
the system.

To see that this might be an imperfect assumption, consider a doped system 
with a small concentration of dopants for
which the Fermi level is pushed into one one of the band tails. Then
the thermal modulation of the atomic coordinates may produce level crossings or
close approaches of the LDA energies. In the doped case, there would normally be
(occupied) levels below the Fermi energy and (unoccupied) levels above. In the
case of a level crossing at the Fermi level, the adiabatic approximation would immediately
transfer charge from one state to another (even if spatially remote), which would induce changes in the dynamics
(since the levels with changed occupation would typically be at least somewhat localized);
and perhaps could lead to a structural change, even a long lived change if the new
structure was ``self-trapping". In practical calculations such charge transfer would not
have be quite so large an affect, since one would usually smear the 
Fermi function slightly from
the $T=0$ step form. The dynamics are then affected by such level crossings or close
approaches even in the adiabatic approximation. A better theory~[21] is due to Allen,
in which interatomic forces are computed from mixed state wave functions obtained by a direct 
integration of the {\it time-dependent}  Schr\"odinger equation (rather than the ``pure"
states computed anew at each new time step for the atomic dynamics in the adiabatic
picture). This approach was formulated to model the dynamical response of a
system to light (in which the simplest model of light-solid
interactions would just involve promotions of electrons to low-lying
conduction states), but the formalism is similar (and much simpler) for the case of
purely phonon-induced level crossings as in our problem. It is possible in fact that the
adiabatic approximation may often be satisfactory for this kind of problem, but the
question has not yet been properly investigated because of difficulties with
post-adiabatic calculations. One case for which nonadiabatic dynamics is {\it essential}
is in the modeling of the thermalization of excited carriers to tail states. This process
has never been realistically modeled despite its manifest importance to a host
of problems in crystalline and non-crystalline semiconductor physics.
More sophisticated methods 
for non-adiabatic dynamics are discussed in the quantum chemistry
literature~[22]; while fundamentally sound, these are currently too difficult
to implement for the large model systems we must consider here.

\vskip .5cm
\noindent{\bf 6. Conclusions}

We have used current computational techniques to demonstrate the nature of the
localized to extended transition in amorphous Si, and then took the additional
step of adding thermal disorder and found that localized states could be time
dependent. A simple analytic model was then presented, which explains some
of the general temperature-dependent features observed in experiments. Finally
we speculated about the importance of non-adiabatic dynamics to these calculations.

I note in closing that the effects reported here are presumably not peculiar to a-Si; they
can be anticipated for any disordered insulator.

\vskip .5truecm
\noindent{\bf Acknowledgements}

This work was supported in part by the National Science Foundation under 
grant DMR-96-18789. I thank my longtime collaborator Prof. Peter Fedders for many
helpful discussions, and Dr. Jianjun Dong for his thesis work on the resonant cluster
proliferation model.
\vskip 0.7truecm
\noindent{\bf REFERENCES}
\begin{itemize}
\item[[1]]
See, for example, P. A. Fedders and D. A. Drabold, Phys. Rev. B {\bf 47} (1993) 13277,
and references therein.
\item[[2]]
P. A. Fedders, D. A. Drabold, P. Ordej\'on, G. Fabricius, D. Sanchez-Portal, E. Artacho
and J. M. Soler (Phys. Rev. B in press, October 15, 1999)
\item[[3]]
F. Wooten and D. Weaire, Solid State Physics, edited by H. Ehrenreich and D. Turnbull 
(Academic Press, New York, 1991), Vol. 40, p.2
\item[[4]]
B. R. Djordjevic, M. F. Thorpe and F. Wooten, Phys. Rev. B {\bf 52} (1995) 5685.
\item[[5]]
D. A. Drabold and P. A. Fedders, Phys. Rev. B {\bf 60} (1999) R721.
\item[[6]]
Jianjun Dong and D. A. Drabold, Phys. Rev. Lett. {\bf 80} (1998) 1928.
\item[[7]]
I. S. Duff, J. K. Reid, N. Munksgaard and H. B. Nielsen, J. Inst. Math. App. {\bf 23} (1979) 235;
M. T. Jones and M. L. Patrick, SIAM J. Matrix Anal. App. {\bf 14} (1993) 553.
\item[[8]]
O. F. Sankey and D. J. Niklewski, Phys. Rev. B {\bf 40} (1989) 3979;
A. A. Demkov, J. Ortega, O. F. Sankey and M. Grumbach, Phys. Rev. B
{\bf 52} (1995) 1618.
\item[[9]]
C. Herring, Phys. Rev. {\bf 57} (1940) 1169.
\item[[10]]
U. Stephan and D. A. Drabold, Phys. Rev. B {\bf 57} (1998) 6391; U. Stephan, R. M. Martin
and D. A. Drabold, {\it Extended range computation of Wannier functions in amorphous
semiconductors}, submitted to Nature.
\item[[11]]
S. Aljishi, J. D. Cohen, and L. Ley, Phys. Rev. Lett. {\bf 64} (1990) 2811.
\item[[12]]
N. F. Mott and E. A. Davis, {\it Electronic processes in non-crystalline materials},
2nd ed. (Clarendon, Oxford, 1979).
\item[[13]]
V. I. Arkhipov and G. J. Adriaenssens, J. Non-Cryst. Sol. {\bf 227} (1998) 166;
Phys. Rev. B {\bf 54} (1996) 16696.
\item[[14]]
D. A. Drabold, P. A. Fedders, S. Klemm and O. F. Sankey, Phys. Rev. Lett. {\bf 
67} (1991) 2179.
\item[[15]]
M. Cobb and D. A. Drabold, Phys. Rev. B {\bf 56} (1997) 3054.
\item[[16]]
For an mpeg format animation, see: http://www.phy.ohiou.edu/$^\sim$drabold/fluctuate.html
\item[[17]]
E. Gibbon, {\it The Decline and Fall of the Roman Empire}, abridged edition by D. M. Low,
(Harcourt Brace, New York, 1960).
\item[[18]]
D. A. Drabold in {\it Insulating and semiconducting glasses} edited by P. Boolchand
(World Scientific, Singapore, 1999).
\item[[19]]
See, for example P. Thomas and S. D. Baranovskii, J. Non-Cryst. Sol. {\bf 164}, (1996) 431
and references therein.
\item[[20]]
R. P. Feynman, Phys. Rev. {\bf 56}(1939) 340; H. Hellmann, {\it Einfuhrung in die
Quantumchemie} (Franz Deutsche, Leipzig, 1937)
\item[[21]]
R. E. Allen, Phys. Rev. B {\bf 50} (1994) 18629.
\item[[22]] See for example, K. Thompson and T. J. Martinez,
J. Chem. Phys. {\bf 110} (1999) 1376 and references therein.

\end{itemize}

\end{document}